# Collaborative Topic Regression with Social Matrix Factorization for Recommendation Systems

**Sanjay Purushotham**[1] SPURUSHO@USC.EDU
**Yan Liu**[2] YANLIU.CS@USC.EDU
**C.-C. Jay Kuo**[1,2] CCKUO@SIPI.USC.EDU
[1]Ming Hsieh Dept. of Electrical Engineering, University of Southern California, Los Angeles, CA 90089 USA
[2]Department of Computer Science, University of Southern California, Los Angeles, CA 90089 USA

## Abstract

Social network websites, such as Facebook, YouTube, Lastfm etc, have become a popular platform for users to connect with each other and share content or opinions. They provide rich information for us to study the influence of user's social circle in their decision process. In this paper, we are interested in examining the effectiveness of social network information to predict the user's ratings of items. We propose a novel hierarchical Bayesian model which jointly incorporates topic modeling and probabilistic matrix factorization of social networks. A major advantage of our model is to automatically infer useful latent topics and social information as well as their importance to collaborative filtering from the training data. Empirical experiments on two large-scale datasets show that our algorithm provides a more effective recommendation system than the state-of-the art approaches. Our results reveal interesting insight that the social circles have more influence on people's decisions about the usefulness of information (e.g., bookmarking preference on Delicious) than personal taste (e.g., music preference on Lastfm). We also examine and discuss solutions on potential information leak in many recommendation systems that utilize social information.

## 1. Introduction

Social networking sites such as Facebook, YouTube, and Lastfm, have become a popular platform for users to connect with friends and share contents (e.g., music, images, and news). The availability of social networks between people have significantly enriched the semantics of links and contents on the web. A fundamental question is whether and how social networks can help to improve recommendation systems, such as products recommendations, advertisement targeting, and scientific paper suggestions. In particular, given the rich content information available, will we have any additional gain by considering social networks? The answer to this question is of great interest to both academia and industries. This paper aims to provide useful insights along this direction.

Collaborative Filtering (CF), which automatically predicts the interests of a particular user based on the collective rating records of similar users or items, has been extensively studied in the literature (Hu et al., 2008; Salakhutdinov & Mnih, 2008; Su & Khoshgoftaar, 2009). The underlying assumption in traditional CF models is that similar users would prefer similar items. However, CF-based models suffer from the sparsity problem and imbalance of rating data, especially for new and infrequent users. Thus, the predicted ratings from CF-models can be unreliable. To overcome the major weaknesses of CF-based recommendation systems, many models have been proposed to explore additional information, such as item's content information (Basilico & Hofmann, 2004; Wang & Blei, 2011) and user's social network (Jamali & Ester, 2009; Ma et al., 2008). For example, Collaborative Topic Regression (CTR) is a state-of-the-art model which naturally incorporates content information via latent dirchelet allocation (Blei et al., 2003) into collaborative filtering framework. Social network-based CF models were recently proposed to find the user's like-minded neigh-bors to address the rating sparsity limitation. As we can see, most existing work have been focused on utilizing either content or social network information, but few have considered them jointly.

In this paper, we propose a hierarchical Bayesian model to integrate social network structure (using matrix factorization) and item content-information (using LDA model) for item recommendation. We connect these two data sources through the shared user latent feature space. The matrix factorization of social network will learn the low-rank user latent feature space, while topic modeling provides a content representation of the items in the item latent feature space, in order to make social recommendations. Our experimental results on two large datasets (*lastfm and delicious* (Cantador et al., 2011)) show that our proposed model outperforms the state-of-the art collaborative filtering-based algorithms such as CTR and Probabilistic Matrix Factorization (PMF). More importantly, our model can provide useful insights into how much social network information can help improve the prediction performance. Furthermore, we introduce the concept of social information leak in recommendation systems and discuss some preliminary yet interesting results. The remainder of this paper is arranged as follows: in section 2, we provide an overview of related works on recommendation systems. In section 3, we present our proposed model and discuss how to learn parameters and do inference. The experimental results and discussion is presented in section 4, followed by conclusions and future work in section 5.

## 2. Related Work

In this section, we review the literature of a few state-of-the art approaches proposed for Collaborative filtering (CF)-based recommendation systems. There are mainly two types of CF-based approaches (1) memory-based approaches, (2) model-based approaches. The memory-based approaches use either user-based approaches (Herlocker et al., 1999) or item-based approaches (Karypis, 2001) for prediction (recommendation) of ratings for items. Even though memory-based approaches are easy to implement and popular; they do not guarantee good prediction results. On the other hand, model-based approaches include several model based learning methods such as clustering models, and the latent factor models. These model-based approaches, especially latent-factor models based on matrix factorization (Koren et al., 2009; Salakhutdinov & Mnih, 2008), have shown promise in better rating prediction since they efficiently incorporate user interests into the model. However, all of the above CF-based approaches assume users are independent and identically distributed and ignore additional information such as the content of item and social connections of users while performing the recommendation task.

Collaborative Topic Regression (CTR) model has been recently proposed (Wang & Blei, 2011), to do collaborative filtering based on probabilistic topic modeling approach. Figure (1) shows the CTR model. The CTR model combines the merits of both traditional collaborative filtering and probabilistic topic modeling approaches. CTR represents users with topic interests and assumes that items (documents) are generated by a topic model. CTR additionally includes a latent variable $\epsilon_j$ which offsets the topic proportions $\theta_j$ when modeling the user ratings. This offset variable $\epsilon$, can capture the item preference of a particular user based on their ratings. Assume there are K topics $\boldsymbol{\beta} = \beta_{1:K}$. The generative process of CTR model is as follows:

1. For each user i, draw user latent vector $u_i \sim \mathcal{N}(0, \lambda_u^{-1} I_K)$
2. For each item j;
   (a) Draw topic proportions $\theta_j \sim$ Dirichlet($\alpha$)
   (b) Draw item latent offset $\epsilon_j \sim \mathcal{N}(0, \lambda_v^{-1} I_K)$ and set the item latent vector as $v_j = \epsilon_j + \theta_j$
   (c) For each word $w_{jn}$,
      i. Draw topic assignment $z_{jn} \sim$ Mult ($\theta$)
      ii. Draw word $w_{jn} \sim$ Mult($\beta_{z_{jn}}$)
3. For each user-item pair (i,j), draw the rating

$$r_{ij} \sim \mathcal{N}(u_i^T v_j, c_{ij}^{-1})$$

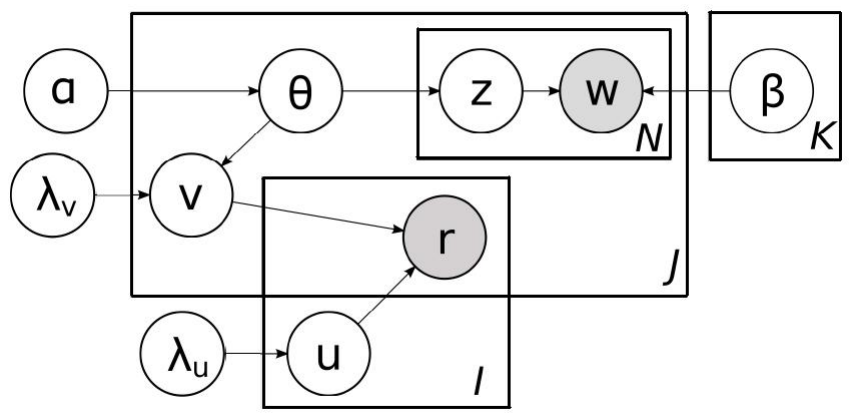

*Figure 1.* Collaborative Topic Regression Model (Wang & Blei, 2011)

CTR model does a good job in using content information for recommendation of items. However, this model does not reliably learn the user latent space for new or inactive users. It has been well-studied and established in social sciences and social network analysis research areas that user's social relations affect user's decision process and their interests. For example: users generally trust their friend's recommendation to buy an item/watch a movie. More recently, recommendation techniques have been developed to incorporate the social relationship information with CF techniques. (Ma et al., 2008) proposed a social recommendation system, based on matrix factorization techniques, which uses user's social network information and user's rating records to recommend products and movies. However, their model cannot be used for recommendation of new or unseen items. In our work, we propose a novel probabilistic model to address the recommendation problem when user's item content, rating records and social network information are all known. Our model can predict ratings for new/unseen items and new/inactive users of a social network.

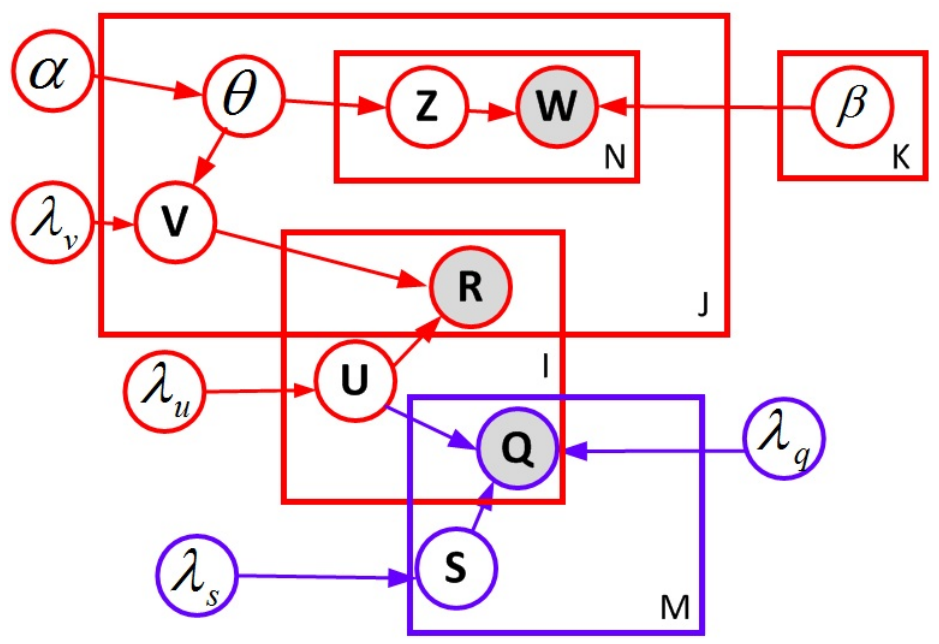

*Figure 2.* Proposed Model - CTR with SMF, CTR part shown in red color, (SMF) Social Matrix Factorization shown in blue color

## 3. Proposed Approach

In this section, we discuss our proposed model, shown in Figure 2. Our model is generalized hierarchical Bayesian model which jointly learns the user, item and social factor latent spaces. We use LDA (Blei et al., 2003) to capture item's content information in latent topic space, and we use matrix factorization to derive latent feature space of user from his social network graph. It can be seen that CTR model (Wang & Blei, 2011) and social matrix factorization (Ma et al., 2008) can be derived as special cases from our proposed model. Our model fuses LDA with social matrix factorization (SMF) to obtain a consistent and compact feature representation. First, we discuss the social matrix factorization and then we will discuss the factorization for our complete model.

Consider a social network graph $\mathcal{G} = (\mathcal{V}, E)$, where the users and their social relations are respectively represented as the vertex set $\mathcal{V} = \{v_i\}_{i=1}^{m}$ and the edge set $E$ of $\mathcal{G}$. Let Q $= q_{ik}$ denote the $m \times m$ matrix of $\mathcal{G}$, which is the social network matrix in this paper. For any pair of vertices $v_i$ and $v_k$, let $q_{ik}$ denote the relation between two users 'i' and 'k'. We associate $q_{ik}$ with a confidence parameter $d_{ik}$, which is used to capture the strength of the user relations. A high value of $d_{ik}$ indicates that user 'i' has a stronger connection (likeliness) with user 'k'. Therefore, the idea of social network matrix factorization is to derive $l$-dimensional feature representation of users, based on analyzing the social network graph $\mathcal{G}$. Let $U \in R^{l \times m}$ and $S \in R^{l \times m}$ be the latent user and social factor feature matrices, with column vectors $U_i$ and $S_k$ representing the user-specific and social factor-specific latent feature vectors respectively. The conditional distribution over the observed social network relationships can be shown as

$$P(Q|U, S, \sigma_Q^2) = \prod_{i=1}^{m} \prod_{k=1}^{m} \mathcal{N}(q_{ij}|g(U_i^T S_k), \sigma_Q^2)^{I_{ij}^Q} \quad (1)$$

where $\mathcal{N}(x|\mu, \sigma^2)$ is the pdf of Gaussian distribution with mean $\mu$ and variance $\sigma_Q^2$, and $I_{ik}^Q$ is the indicator function that is '1' if user 'i' and user 'k' are connected in the social graph (i.e. there is an edge between the vertices 'i' and 'k'), and equal to 0 otherwise. The function $g(x)$ is the logistic function $g(x) = \frac{1}{1+exp(-x)}$, which bounds the range of $U_i^T S_k$ within $[0, 1]$. We place zero-mean spherical Gaussian priors on user and factor feature vectors:

$$P(U|\sigma_U^2) = \prod_{i=1}^{m} \mathcal{N}(U_i|0, \sigma_U^2 I) \quad (2)$$

$$P(S|\sigma_S^2) = \prod_{k=1}^{m} \mathcal{N}(S_k|0, \sigma_S^2 I) \quad (3)$$

Hence, through Bayesian inference, we have

$$p(U, S|Q, \sigma_Q^2, \sigma_U^2, \sigma_S^2) \propto p(Q|U, S, \sigma_Q^2) p(U|\sigma_U^2) p(S|\sigma_S^2)$$

Now, combining LDA with SMF (figure 2), we have

$$\begin{aligned} &p(U, V, S|Q, R, \sigma_Q^2, \sigma_R^2, \sigma_U^2, \sigma_V^2, \sigma_S^2) \\ &\propto p(R|U, V, \sigma_R^2) p(Q|U, S, \sigma_Q^2) \\ &\times p(U|\sigma_U^2) p(V|\sigma_V^2) p(S|\sigma_S^2). \end{aligned} \quad (4)$$

The log of the posterior distribution for the above equation can be found by substituting the corresponding pdfs. Note: the item latent vector $v_j$ is generated by a key property due to CTR.

$$P(V|\sigma_V^2) \sim \mathcal{N}(\theta_j, \lambda_V^{-1} I_k) \quad (5)$$

where $\lambda_V = \sigma_R^2 / \sigma_V^2$.

### 3.1. Learning the parameters of our model

For learning the parameters, we develop an EM-style algorithm similar to (Wang & Blei, 2011). Maximization of the posterior is equivalent to maximizing the complete log-likelihood of $U, V, S, \theta_{1:J}$,$R$ and $Q$ given $\lambda_U, \lambda_V, \lambda_S, \lambda_Q$ and $\beta$.

$$\begin{aligned} \mathcal{L} = &-\frac{\lambda_U}{2} \sum_i u_i^T u_i - \frac{\lambda_V}{2} \sum_j (v_j - \theta_j)^T (v_j - \theta_j) \\ &+ \sum_j \sum_n log\left( \sum_k \theta_{jk} \beta_{k, w_{jn}} \right) - \sum_{ij} \frac{c_{ij}}{2} (r_{ij} - u_i^T v_j)^2 \\ &- \frac{\lambda_Q}{2} \sum_{i,m} \frac{d_{im}}{2} (q_{im} - u_i^T s_m)^2 - \frac{\lambda_S}{2} \sum_k s_k^T s_k \end{aligned} \quad (6)$$

where $\lambda_U = \sigma_R^2/\sigma_U^2, \lambda_S = \sigma_R^2/\sigma_S^2, \lambda_Q = \sigma_R^2/\sigma_Q^2$ and Dirichlet prior ($\alpha$) is set to 1. We optimize this function by gradient ascent approach by iteratively optimizing the collaborative filtering and social network variables $u_i, v_j, s_m$ and topic proportions $\theta_j$. For $u_i, v_j, s_m$, maximization follows similar to matrix factorization (Hu et al., 2008). Given a current estimate of $\theta_j$, taking the gradient of $\mathcal{L}$ with respect to $u_i, v_j$ and $s_m$ and setting it to zero helps to find $u_i, v_j, s_m$ in terms of $U, V, C, R, S, \lambda_V, \lambda_U, \lambda_S, \lambda_Q$. Solving the corresponding equations will lead to the following update equations:

$$u_i \leftarrow (VC_iV^T + \lambda_Q SD_iS^T + \lambda_U I_K)^{-1} (VC_iR_i + \lambda_Q SD_iQ_i) \quad (7)$$

$$v_j \leftarrow (UC_jU^T + \lambda_V I_K)^{-1}(UC_jR_j + \lambda_V \theta_j) \quad (8)$$

$$s_m \leftarrow (\lambda_Q UD_mU^T + \lambda_S I_K)^{-1}(\lambda_Q UD_mQ_m) \quad (9)$$

where $C_i, D_i$ are diagonal matrices with $c_{ij}, d_{ij}; j = 1....J$ as its diagonal elements and $R_i = (r_{ij})_{j=1}^J$ for user $i$. For each item $j, C_j$ and $R_j$ are similarly defined. Note that $c_{ij}$ is confidence parameter for rating $r_{ij}$, for more details refer (Wang & Blei, 2011). We define $d_{ij}$ as the confidence parameter for $q_{ik}$, where $q_{ik}$ is the relationship between users $i$ and $k$. The equation (8) shows how topic proportions $\theta_j$ affects the item latent vector $v_j$, where $\lambda_V$ balances this effect. Given U and V, we can learn the topic proportions $\theta_j$. We define $q(z_{jn} = k) = \phi_{jnk}$ and then we separate the items that contain $\theta_j$ and apply Jensen's inequality:

$$\mathcal{L}(\theta_j) \geq -\frac{\lambda_V}{2}(v_j - \theta_j)^T(v_j - \theta_j) + \sum_n \sum_k \phi_{jnk}(\log \theta_{jk}\beta_{k,w_{jn}} - \log \phi_{jnk}) = \mathcal{L}(\theta_j, \phi_j) \quad (10)$$

The optimal $\phi_{ink}$ satisfies $\phi_{jnk} \propto \theta_{jk}\beta_{k,w_{jn}}$. Note, we cannot optimize $\theta_j$ analytically, so we use projection gradient approaches to optimize $\boldsymbol{\theta_{1:J}}$ and other parameters $U, V, \boldsymbol{\phi_{1:J}}$. After we estimate U,V and $\boldsymbol{\phi}$, we can optimize $\beta$,

$$\beta_{kw} \propto \sum_j \sum_n \phi_{jnk} 1[w_{jn} = w] \quad (11)$$

### 3.2. Prediction

After the optimal parameters $U^*, V^*, \theta_{1:J}^*$ and $\beta^*$ are learned, our proposed model can be used for in-matrix and out-matrix prediction (recommendation) tasks. If $D$ is the observed data, then both in-matrix and out-matrix predictions can be easily estimated. As discussed in (Wang & Blei, 2011), in-matrix prediction refers to the case where the user has not rated an item but that item has been rated by atleast one other user. On the other hand, out-matrix refers to the case where none of the users have rated a particular item i.e. the item has no rating records. For in-matrix prediction, we use the point estimate of $u, \theta_j$ and $\epsilon_j$ to approximate their expectations as:

$$\mathcal{E}[r_{ij}|D] \approx \mathcal{E}[u_i|D]^T(\mathcal{E}[\theta_j|D] + \mathcal{E}[\epsilon_j|D]) \quad (12)$$

$$r_{ij}^* \approx (u_i^*)^T v_j^* \quad (13)$$

For out-matrix prediction, the item is new and has not been rated by other users. Thus, $\mathcal{E}[\epsilon_j] = 0$ and we predict the ratings as:

$$\mathcal{E}[r_{ij}|D] \approx \mathcal{E}[u_i|D]^T(\mathcal{E}[\theta_j|D]) \quad (14)$$

$$r_{ij}^* \approx (u_i^*)^T \theta_j^* \quad (15)$$

## 4. Experimental Analysis

We conduct several experiments to compare the performance of our proposed model with the state-of-the-art techniques. We evaluate our model on real-world datasets for music and bookmark recommendations. Our experiments help us to answer two key questions:

- How does our model compare with respect to the state-of-the-art collaborative filtering techniques?
- How does content parameter $\lambda_v$ and social network parameter $\lambda_q$ affect the prediction accuracy?

### 4.1. Description of Datasets

Table 1 shows the description of two real-world datasets considered for our experiments: *hetrec2011-lastfm-2k* (Lastfm) and *hetrec2011-delicious-2k* (Delicious) (Cantador et al., 2011). The datasets are first preprocessed to remove noisy entries. For *hetrec2011-lastfm-2k* dataset, if the user has listened to an artist (item) then we consider the user rating for the artist as '1', else we do not give any user ratings for the artist. Similarly, for *hetrec2011-delicious-2k* dataset, if the user has bookmarked an URL (item) then we consider the rating for that bookmarked URL as '1', else we do not give any user ratings for the URL. We consider artists and URLs as items in the above two datasets. We observe that the user-item matrices for both the datasets are highly sparse (99.7% sparse and 99.91% sparse respectively).

*Table 1.* Dataset description

| Dataset | Lastfm | Delicious |
|---|---|---|
| users | 1892 | 1867 |
| items | 17632 | 69226 |
| tags | 11946 | 53388 |
| user-user relations | 25434 | 15328 |
| user-tags-items | 186479 | 437593 |
| user-items relations | 92834 | 104799 |

### 4.2. Evaluation

In our experiments, we split each of the datasets into two parts - training (90%) and testing datasets (10%). The model is trained on a majority of training dataset and the optimal parameters are obtained on a small held-out dataset. Using the optimal parameters, the ratings are predicted for the entries in the testing datasets. For evaluation, we consider 'recall' as our performance evaluation metric, since 'precision' metric is difficult to evaluate (zero rating for item can imply either the user does not like the item or does not know about the item). Recall only considers the rated items within the top M - a higher recall with lower M implies a better system. For each user, we define the $recall@M$ as:

$$recall@M = \frac{\text{number of items the user likes in Top M}}{\text{total number of items the user likes}}$$

The above equation calculates user-oriented recall. We can similarly define item-oriented recall. For consistency and convenience, we use user-oriented recall for in-matrix prediction throughout this paper.

### 4.3. Experimental settings

For collaborative filtering based on matrix factorization(denoted by CF), we used grid search to find the parameters such that we get good performance on the testing dataset. We found that $\lambda_v = 100, \lambda_u = 0.01, a = 1, b = 0.01, K = 200$ gives good performance for CF approach. Note: $a$ and $b$ are tuning parameters ($a > b > 0$) for the confidence parameters $c_{ij}$ and $d_{ij}$ (equation (6)). For Collaborative Topic Regression model (denoted by CTR), we choose the parameters similar to CF approach. We set the parameters $\lambda_u = 0.01, a = 1, b = 0.01, K = 200$ and we vary the parameter $\lambda_v$ to study its effect on the prediction accuracy. For our model, we set parameters $a = 1, b = 0.01$, and vary all other parameters to study their affect on prediction accuracy. We use the terms prediction accuracy and recall interchangeably throughout this paper.

### 4.4. Comparisons

We compare our proposed model with some of the state-of-the-art algorithms such as Collaborative Topic Regression (Wang & Blei, 2011), and Matrix Factorization (Koren et al., 2009). First, we study the effect of precision parameter $\lambda_v$ on the CTR model. Figure 3 shows that when $\lambda_v$ is small in CTR, the per-item latent vector $v_j$ can diverge significantly from the topic proportions $\theta_j$ which was observed in (Wang & Blei, 2011). Figure 4 shows that in-fact we observe

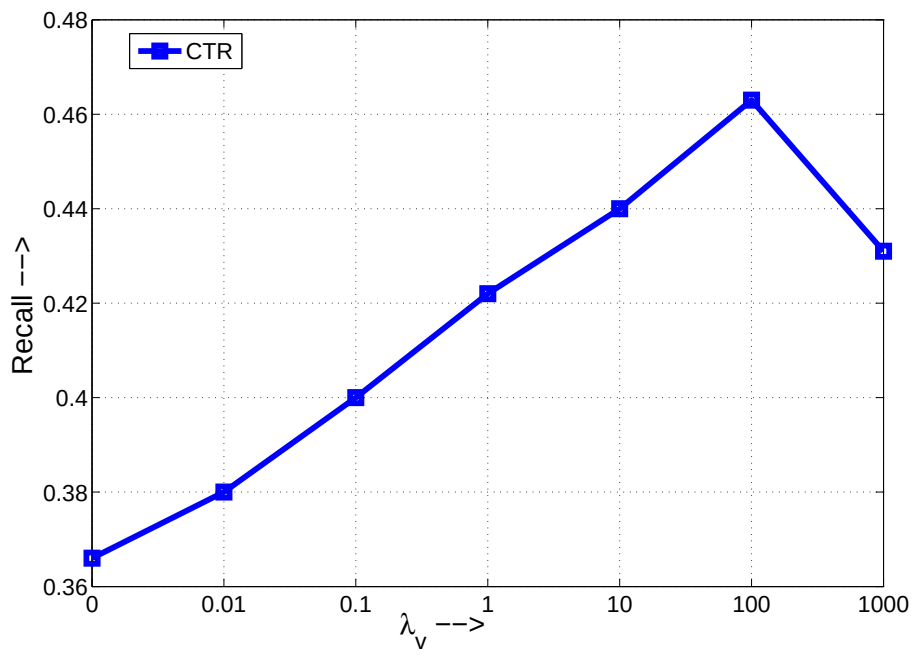

*Figure 3.* Recall of in-matrix prediction task for CTR model by varying content parameter $\lambda_v$ and fixing number of recommended items i.e $M = 250$. Dataset used: *hetrec2011-lastfm-2k*

the same e ffect of $\lambda_v$ for our model. Moreover, the plots in figure 4 clearly show that our proposed model consistently outperforms CTR model by a margin of 2.5 $\sim$ 3% for both the datasets. This can be explained since our model uses social network information to better model the user latent space i.e. it better models user's preferences from similar friends. Figure 5 shows the overall performance for in-matrix prediction when we vary the number of returned items $M = 50, 100, ..., 250$ while keeping $\lambda_v (= 100)$ as constant. This plot shows that as the number of returned items is increased, the performance of our model improves. This figure also shows that our approach always outperforms both CTR and CF approaches at different values of M. We observed that the recall measured at smaller M ($M < 50$) is quite small for all the models, since the average number of items per user in the test dataset is quite small and so the models recommend most popular items in top M.

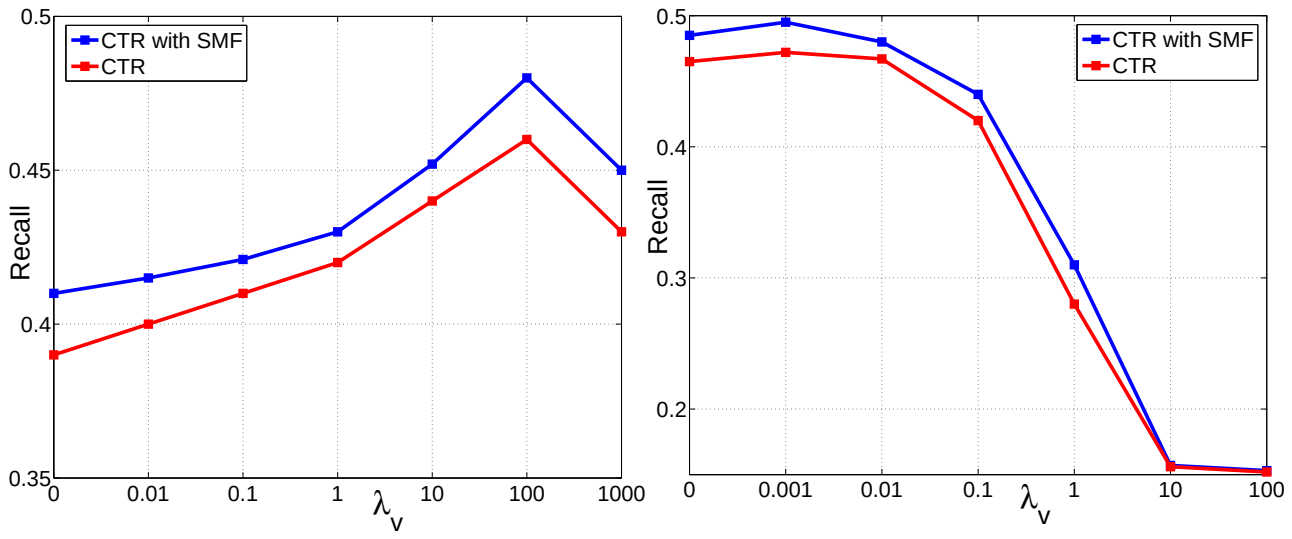

*Figure 4.* Comparison of Recall for CTR and our proposed model (CTR with SMF) by varying $\lambda_v$ and fixing $M = 250$. Left plot: Dataset used is *hetrec2011-lastfm-2k*, Right plot: Dataset used is *hetrec2011-delicious-2k*

### 4.5. Impact of parameters $\lambda_v$, $\lambda_q$

Our model allows us to study how the content ($\lambda_v$) and social network parameters ($\lambda_q$) affect the overall performance of the recommendation system. Here we discuss how to balance these parameters to achieve better recommendation of items. If $\lambda_q = 0$, our model

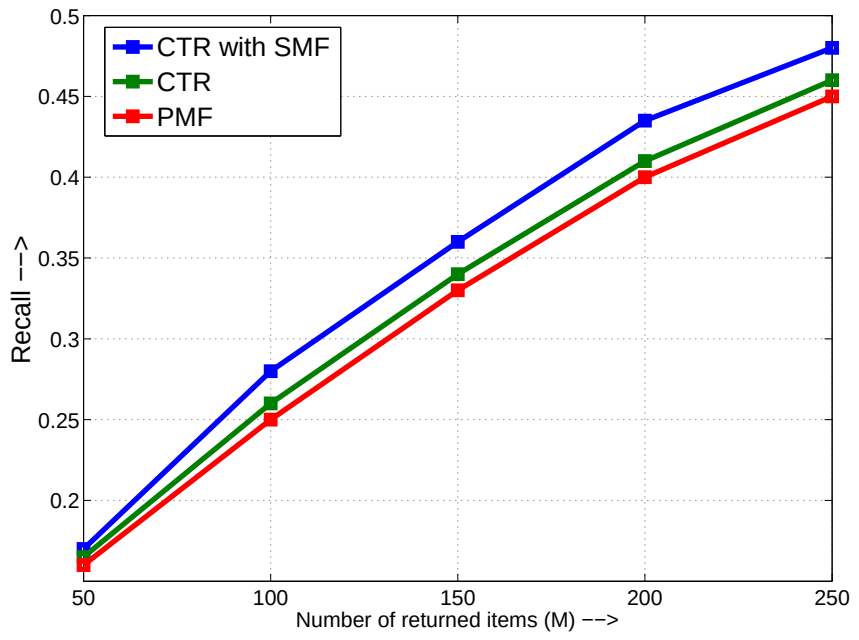

*Figure 5.* Recall comparison of various models for in-matrix prediction task by varying number of recommended items $M$ and fixing $\lambda_v = 100$. Dataset used: *hetrec2011-lastfm-2k*. Our model indicated by CTR with SMF. PMF indicates matrix factorization (CF)

collapses to the CTR model which uses topic modeling and user-item rating matrix for prediction. If $\lambda_q = \infty$, our model uses only the information from social network to model user's preferences. In all other cases, our model fuses information from topic model and the user social network for matrix factorization and furthermore to predict the ratings for users. Figure 6 shows how our system performs when the social network parameter $\lambda_q$ is varied while keeping the content parameter $\lambda_v$ as constant (fixed topic model). From this figure, we observe that the value of $\lambda_q$ impacts the recommendation results significantly, which demonstrates that fusing user social network with topic model (CTR) improves recommendation accuracy considerably (2.5 $\sim$3%). Figure 6 also indicates that for small values of $\lambda_q$, the improvement in prediction accuracy is small and negligible, and it increases with further increase in $\lambda_q$. However, when $\lambda_q$ increases beyond a certain threshold, the prediction accuracy decreases with further increase in $\lambda_q$. This can be intuitively explained as follows: for large values of $\lambda_q$, our model gives more preference to social network information (similar neighbor) but less preference to user's tastes (previous item rating records) and hence the prediction accuracy may not be reliable for large $\lambda_q$. Our model achieves best prediction accuracy for *hetrec2011-lastfm-2k* dataset around $\lambda_q \in (100, 200)$ irrespective of the parameter $\lambda_v$'s value. This insensitivity of the optimal values for the parameter $\lambda_q$ shows that our model can be easily trained on the held-out dataset.

To study how the content and social network parameters balance/influence our model's predictions, we plot the recall by varying these parameters. Figure 7 shows the contour and 3D plots of recall for our proposed model when tested on *hetrec2011-lastfm-2k* dataset. When the parameters are zero, i.e $\lambda_v = 0$ and $\lambda_q = 0$, our model reduces to standard CF model - which has a poor prediction accuracy and this is confirmed in these plots. When we increase $\lambda_v$ at fixed $\lambda_q$, we see that our model's performance improves (smaller $\lambda_v$ implies model behaves like CF model). Similar observations can be made for varying $\lambda_q$ at fixed $\lambda_v$. Moreover, there is a region of values for $\lambda_v$ and $\lambda_q$ (near $\sim$ (100,100)), around which our model provides the best performance in terms of recall. To investigate further, we plotted the recall contours for our model around the empirically chosen range $(100, 250)$. Figure 8 show this plot. From this figure, we can infer that there is a region where the optimal values for $\lambda_v$ and $\lambda_q$ ensures the best prediction accuracy for our model. To our surprise, we found that both the parameters had similar value of $\sim$150. To find out if having similar and higher values of $\lambda_q$ and $\lambda_v$ parameters always guarantee best performance for our model, we conducted similar experiments on *hetrec2011-delicious-2k* dataset.

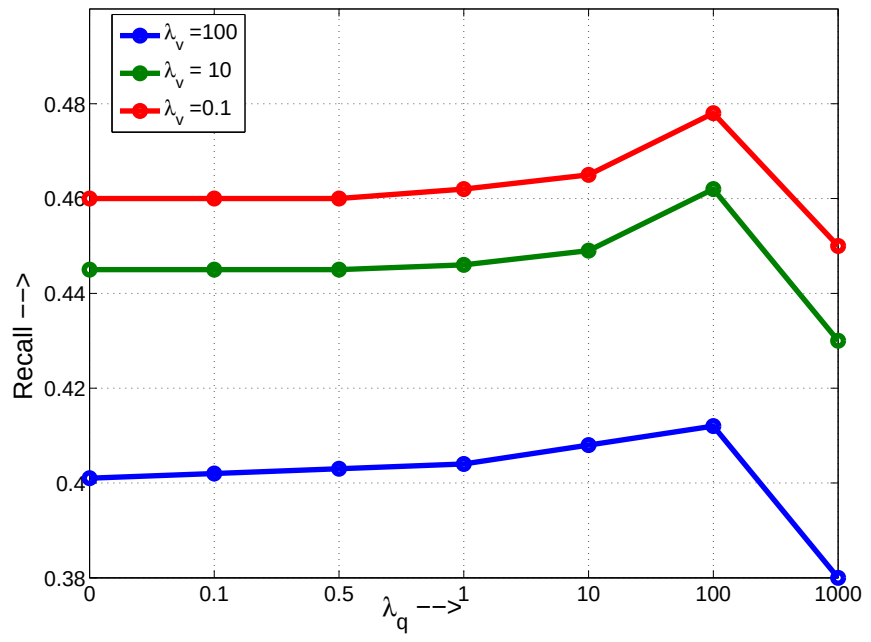

*Figure 6.* Recall of in-matrix prediction task for our proposed model by varying content parameter $\lambda_v$ and social network parameter $\lambda_q$ at $M = 250$. Dataset used: *hetrec2011-lastfm-2k*

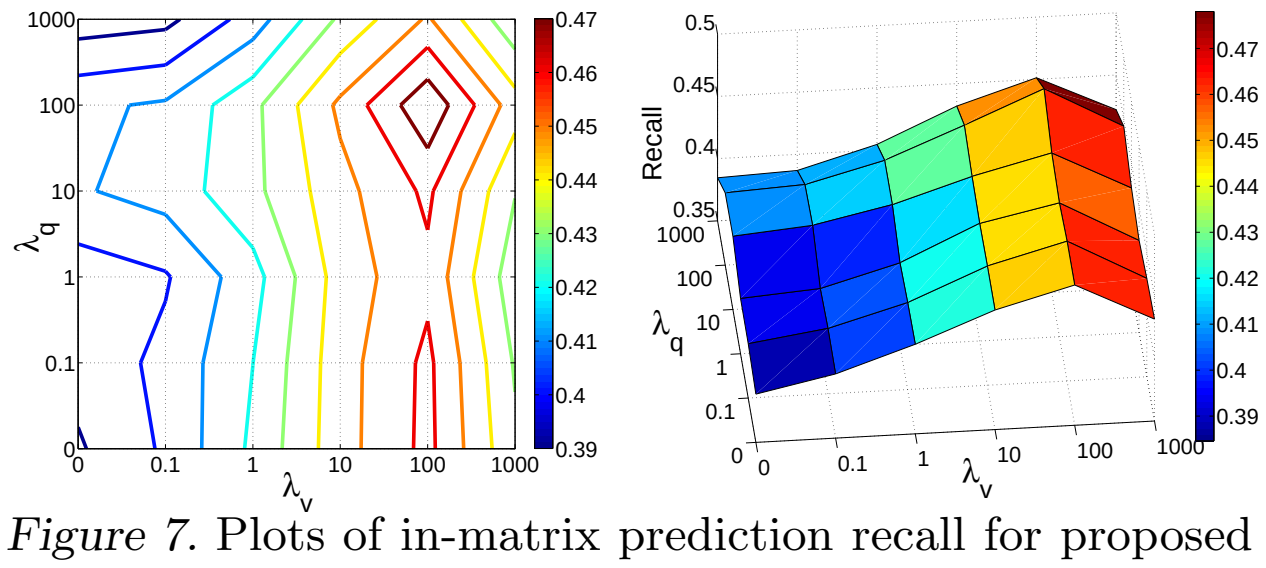

*Figure 7.* Plots of in-matrix prediction recall for proposed model by varying content parameter $\lambda_v$ and social network parameter $\lambda_q$ at $M = 250$. Dataset used: *hetrec2011-lastfm-2k*

Figure 9 shows the contour and 3D-plots of recall by varying $\lambda_v$ and $\lambda_q$ for *hetrec2011-delicious-2k* dataset. We observe that in figure 9 the optimal values for $\lambda_q, \lambda_v$ are pretty small, and our model achieves best prediction accuracy for $\lambda_q = 0.05$ and $\lambda_v = 0.01$. We can explain this by looking at the dataset. *hetrec2011-delicious-2k* is obtained from Delicious social bookmarking website, where majority of user's bookmarks (items) are publicly shared online or with friends.

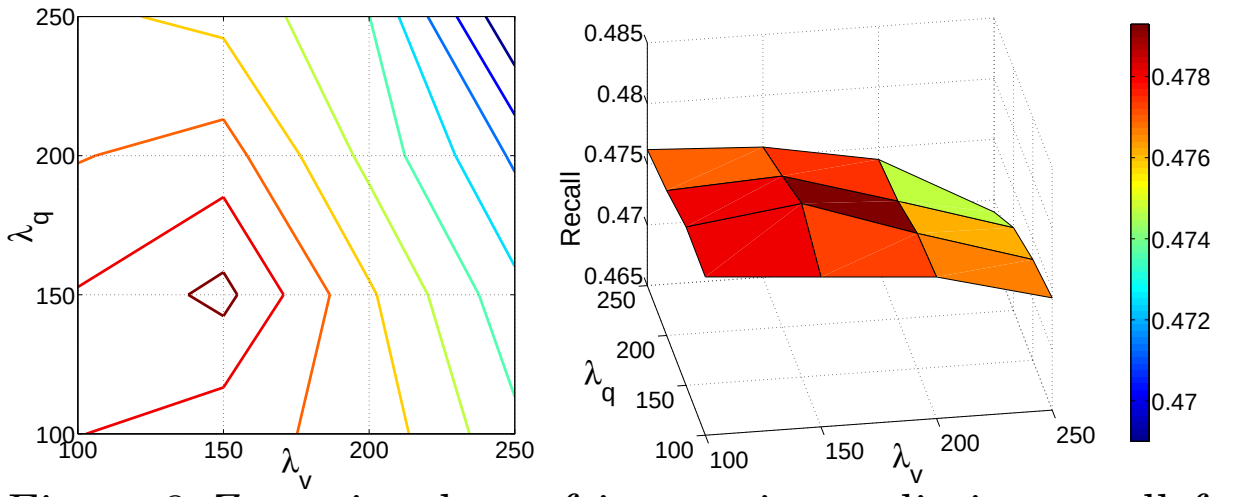

*Figure 8.* Zoom-in plots of in-matrix prediction recall for proposed model by varying content parameter $\lambda_v$ and social network parameter $\lambda_q$ at $M = 250$. Dataset used: *hetrec2011-lastfm-2k*

Thus, the user's social network plays a more important role than the content information of URLs (items) for the URL prediction task. Moreover, since the dataset is highly sparse (99.91%), the content and social network information help for some users only and thus these parameters have smaller values. On the other hand, for *hetrec2011-lastfm-2k* dataset (99.7% sparse) we see that it's a music dataset, and generally the artist's music (item content) has a great influence on user's tastes. Moreover, *lastfm* users tend to be friends with other users who have similar music interests. Thus, we observe that, higher values of parameters $\lambda_v$ and $\lambda_q$ achieves the best prediction for our model. From these experiments (figures 8 and 9), we can say that the optimal values of $\lambda_q$ and $\lambda_v$ are highly dependent on datasets and their values balance how the content and social network information could be used for achieving best prediction accuracy (best recommendation).

### 4.6. Complexity Analysis

Our model utilizes LDA for topic modeling, thus, the time complexity of our model is quite expensive when compared to the traditional matrix factorization techniques. Table 2 shows the average time consumed by our model when compared with CTR model. For *hetrec2011-lastfm-2k* dataset, we observed that when $\lambda_v$ is small, our model converges faster than CTR model, on the other hand when $\lambda_v$ is large, then our model takes more time for convergence. This is because, for smaller values of $\lambda_v$, the update for $v_j$ is done much faster in our model (due to joint learning of latent space vectors). When averaged over all values of $\lambda_v$ our model took comparable time as CTR model. Table 3 shows how our model performs when latent space dimensions ($K$) is varied. We observe that when we use a smaller value for $K$, the accuracy of our model decreases but it converges much faster (45x). This shows that, clearly there is a trade-off between prediction accuracy and $K$(number of topics). Moreover, we observed that using smaller $K$ in our model achieves similar accuracy as the CTR model which uses larger $K$. All our experiments were run on single core processors with 2∼4 GB RAM.

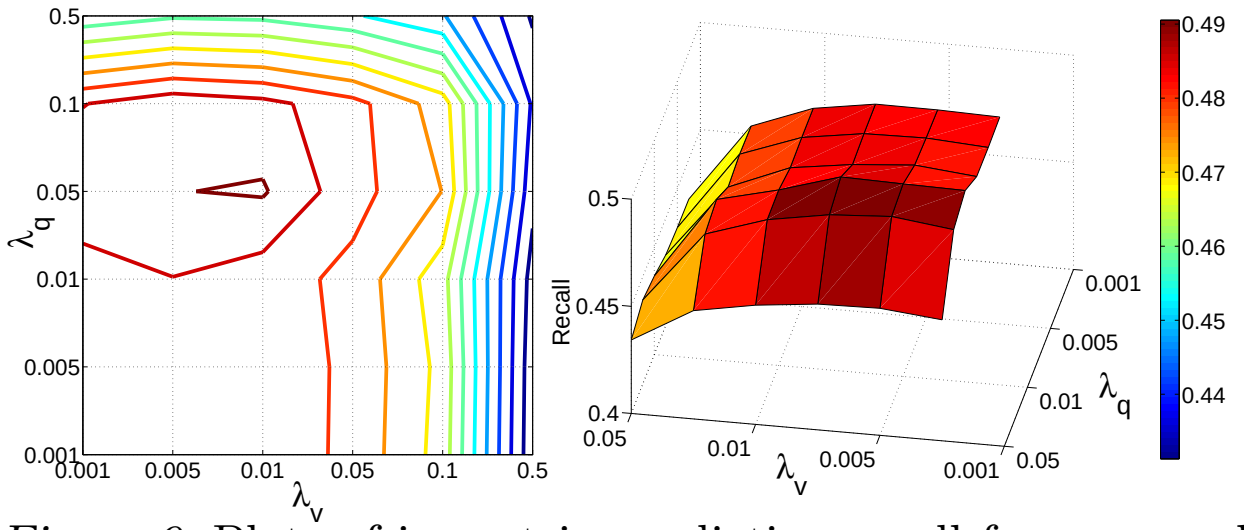

*Figure 9.* Plots of in-matrix prediction recall for proposed model by varying content parameter $\lambda_v$ and social network parameter $\lambda_q$ by fixing $M = 250$. Dataset used: *hetrec2011-delicious-2k*

*Table 2.* Time complexity comparison of our model with respect to CTR model, Dataset:*hetrec2011-lastfm-2k*

| Model | Time taken(hrs) $\lambda_v < 1$ | Time taken(hrs) $\lambda_v > 1$ | Avg. time taken(hrs) |
|---|---|---|---|
| CTR | 9.45 | 9.47 | 9.46 |
| Our model | 8.47 | 10.59 | 9.53 |

*Table 3.* Time complexity of our model for varying latent space dimensions, Dataset used: *hetrec2011-delicious-2k*

| K | Time taken | Avg. recall at $\lambda_v = 0.1, \lambda_q = 0.1$ |
|---|---|---|
| 50 | ∼14.3 mins | 0.472 |
| 200 | ∼10.5 hours | 0.510 |

### 4.7. Discussion on Social Network Structure

In our experiments, we considered a 'final' static social network where the relations of users is fixed (that is, it doesn't change with time). We showed that given the user's social network, our model can more accurately predict the user ratings. It is possible that users form social network because they like similar types of items (music or bookmarks) and this social network dynamically evolves over time. Hence, we feel that using a final static social network could be a source of potential information leak. That is, our model could be making better predictions using future social network information. From our experience, we find that none of the current literature discuss or incorporate this into their recommendation framework.

To test our hypothesis, we conducted new experiments on the *delicious* dataset by considering the evolving social network structure. First, we obtained different training datasets based on the timestamp of the social network, then we evaluated the in-matrix recall on the test users by considering both the 'final' and 'timestamped' social network information. Figure 10 shows that, using 'final' static social network provides better recall then using the timestamped (evolving) social network. In-fact, we observed that smaller the training dataset, the larger the information leak in the recommendation system. We observe that our model obtains 3-5% improvement in prediction accuracy by using full (static) social network as compared to using timestamped (evolving) social network.

## 5. Conclusions and Future Work

In this paper, we presented a generalized hierarchical Bayesian model that exploits that user's social network information and item's content information to recommend items to users. Our experiments on two large social media datasets showed that our proposed model consistently outperforms the state-of-the-art approaches such as CTR and matrix factorization techniques. The main contributions of our paper include: 1) demonstrating the effectiveness of social network to improve the performance of recommendation systems, 2) providing effective trade-off techniques to improve prediction accuracy when both social and content information are available in recommendation system, 3) starting discussion on a new problem (social information leak) in social recommendation systems, which has not been explored in existing literature.

For future work, we are interested in examining parallel implementations of our algorithms so that they are scalable to large-scale datasets; we are also interested in examining how to capture the dynamics of the evolving social network into our model and analyze it's affect (social information leak) on prediction accuracy.

## Acknowledgments

This research is continuing through participation in the Anomaly Detection at Multiple Scales (ADAMS) program sponsored by the U.S. Defense Advanced Research Projects Agency (DARPA) under Agreement Number W911NF-11-C-0200. We thank all the reviewers for their useful comments.

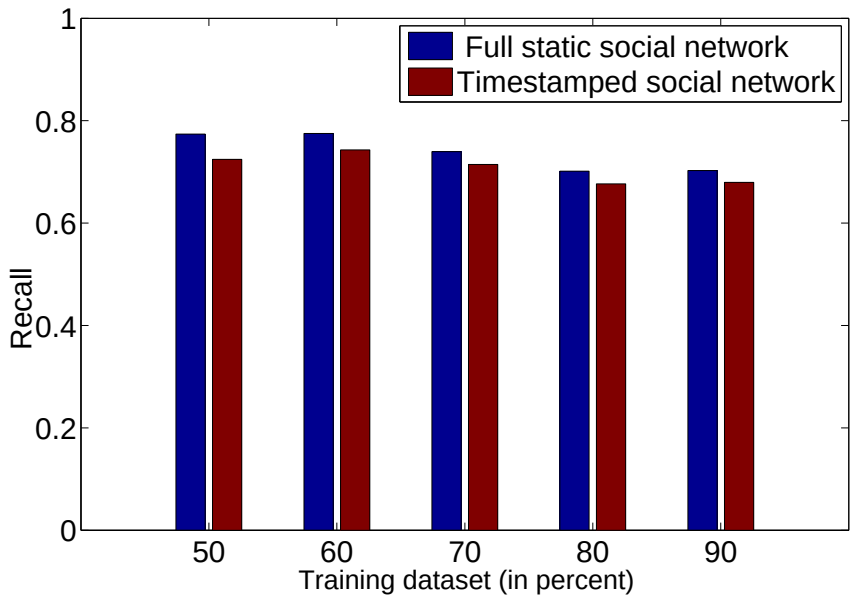

*Figure 10.* Recall of proposed model by varying social network structure. Dataset used: *hetrec2011-delicious-2k*